\documentclass[a4paper,11pt]{article}

\usepackage{pos}
\usepackage{subcaption}
\usepackage{wrapfig}

\title{Towards a cosmic ray composition measurement with the IceAct telescopes at the IceCube Neutrino Observatory}

\ShortTitle{IceAct composition measurement}

\author{The IceCube Collaboration \\{\normalsize \normalfont(a complete list of authors can be found at the end of the proceedings)}\\}

\emailAdd{larissa.paul@icecube.wisc.edu}

\abstract{

The IceCube Neutrino Observatory is equipped with the unique possibility to measure cosmic ray induced air showers simultaneously by their particle footprint on the surface with the IceTop detector and by the high-energy muonic shower component at a depth of more than 1.5 km. Since 2019 additionally two Imaging Air Cherenkov Telescopes, called IceAct, measure the electromagnetic component of air showers in the atmosphere above the IceCube detector. This opens the possibility to measure air shower parameters in three independent detectors and allows to improve mass composition studies with the IceCube data. One IceAct camera consists of 61 SiPM pixels in a hexagonal grid. Each pixel has a field of view of 1.5 degree resulting in an approximately 12-degree field of view per camera. A single telescope tube has a diameter of 50 cm, is built robust enough to withstand the harsh Antarctic conditions, and is able to detect cosmic ray particles with energies above approximately 10 TeV. A Graph Neural Network (GNN) is trained to determine the air shower properties from IceAct data. The composition analysis is then performed using Random Forest Regression (RF). Since all three detectors have a different energy threshold, we train several RFs with different inputs, combining the different detectors and taking advantage of the lower energy threshold of the IceAct telescopes. This will result in composition measurements for different detector combinations and enables cross-checks of the results in overlapping energy bands. We present the method, parameters for data selection, and the status of this analysis.

\vspace{4mm}
{\bfseries Corresponding authors:}
Larissa Paul$^{1,2*}$\\
{$^{1}$ \itshape Marquette University, Milwaukee, Wisconsin}\\
{$^{2}$ \itshape South Dakota School of Mines and Technology, Rapid City, South Dakota}\\\\[4mm]
$^*$ Presenter

\ConferenceLogo{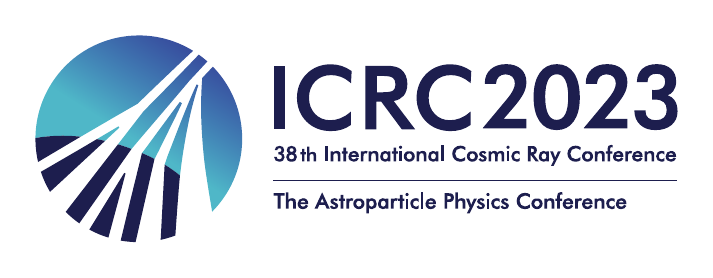}

\FullConference{The 38th International Cosmic Ray Conference (ICRC2023)\\ 26 July -- 3 August, 2023\\ Nagoya, Japan}
}

\begin{document}

\newcommand{\logEGeV}{$\log_{10}(\frac{E}{GeV})$}
\newcommand{\todo}[1]{{\bf \textcolor{red}{#1}}}

\maketitle

\section{Introduction}\label{sec1}
The two IceAct telescopes have been continuously taking data since 2019 during the Austral winter. One -- the "roof telescope" -- is positioned in the center of the IceCube Neutrino Observatory (IceCube)~\cite{Aartsen:2016nxy, icetop2013} on top of the IceCube Laboratory building. The other -- the "field telescope" -- is roughly 220$\,$m west from the roof telescope. 
The telescopes measure the Cherenkov light which is emitted by the high energy particles of air showers produced in the collision of cosmic ray particles with the Earth's atmosphere. The photons collected by the telescopes are mainly emitted by the electromagnetic part of the extensive air shower. This is a complementary measurement to the standard detectors of IceCube which measure the high energy muon component in the deep ice and the electromagnetic plus low energy muon footprint at the surface with the IceTop detector~\cite{icetop2013}. 

The idea of this analysis is to use a Graph Neural Network (GNN) to reconstruct air shower parameters, described in section \ref{secgnn}. The results of the GNN reconstruction are then used for a mass composition analysis performed with a Random Forest Regression (RF) described in section \ref{secrf}. The input parameters into the RF include the air shower parameters 
and the parameters of the IceCube and IceTop detectors. 
In this proceeding, two mass composition studies will be discussed: one using only input parameters from the IceAct telescopes, and the other one using input parameters from the IceAct telescope and the IceCube detector. Each analysis has their own energy range due to the different detector thresholds. For the final analysis step, a template analysis similar to the previously published three year mass composition analysis from IceCube~\cite{IceCube:2019hmk} is performed, which is described in section \ref{sec5}.
\section{Data preparation}\label{secdata}
In 2020, the data acquisition system (DAQ) of both telescopes was unified using the TARGET DAQ which was developed for the Cherenkov Telescope Array \cite{Author:2017Funktarget}. In preparation for the analysis, both telescopes have been calibrated. A detailed description of the calibration runs for the telescopes can be found in \cite{Author:2023icrcLars}. The muon runs and dark count run data described there are used for the calibration of each individual pixel. To perform the calibration, a peak extraction is applied to the data. The peak extraction subtracts the median of the waveform and calculates the parabola through the three highest points of the waveform. If the parabola is pointing upwards the highest point of the parabola is accepted as peak height and peak time. The width of the peak is determined by calculating the time difference between the highest point and the point where the peak reaches half the peak height. 
The applied quality cuts are that the pulse width needs to be at least 2$\,$ns and the peak time needs to be within a time window around the trigger time.
The calibration results in an average gain of 7.2$\,$ADC counts per photon.

The typical data-taking period for the telescopes is from the end of April to the beginning of September, which results in a detector uptime of above 30\% of a full year. The data taking is fully automated and the threshold is regularly adjusted for the amount of background light present, accounting for aurora flares and full moon periods. Threshold scans, calibration data taking, and unfavorable weather conditions such as clouds and full moon periods, reduce the good-run uptime to slightly above 20\% of the year since 2021 \cite{Author:paul2021hybrid}. 
Quality parameters
are currently being studied, for the selection of stable atmospheric conditions. For this, a commercial fish-eye camera has been installed on top of the IceCube Laboratory, which regularly takes pictures of the night sky above the telescopes. This allows for monitoring the night sky background (NSB) e.g.~aurora activities and cloud coverage by counting the number of stars visible in the camera image. Additionally, monitoring parameters of the telescope itself can be used for the data selection as described in \cite{Author:2023icrcLars}.

\section{Simulation dataset}\label{secsim}
For this study, a dedicated Monte Carlo (MC) data set has been produced. A special module has been implemented into the IceCube software framework icetray \cite{icetray}, to optimize the production of the air shower simulations. The module throws randomly the shower core position in IceCube coordinates and calculates the relative telescope positions in CORSIKA \cite{Author:Heck1998vt} coordinates which reduces the data volume by only storing Cherenkov photons around the telescope positions. The events are thrown in a circle centered on the point between the roof telescope and the field telescope, with a radius which increases with increasing energy: 250$\,$m for events below 10$^4\,$GeV, and increasing by 50$\,$m for each 0.25 in \logEGeV{}. The events are drawn from an $E^{-1}$ spectrum between 10$^{3.5}\,$GeV and 10$^{6.75}\,$GeV. To boost the number of low energy events, several additional data sets have been created with smaller energy ranges, all following the $E^{-1}$ spectrum. Six different types of nuclei are used as primary particles in the simulation: proton, helium~(He), nitrogen~(CNO), neon, aluminum~(MgSi), and iron. 
The air shower propagation is simulated using CORSIKA with Sibyll2.3c as the interaction model for the high energy interactions and FLUKA for the low energy hadronic interactions. The atmosphere for the CORSIKA simulation uses a layered parameterization of the average South Pole atmosphere in April \cite{sam_thesis}.
The simulation is then processed to include the response of the IceTop and in-ice detectors. 
The Cherenkov photons produced in the atmosphere by CORSIKA are not absorbed or scattered during the simulation. To account for this, an additional module was implemented which calculates the total transmission probability for each photon depending on the production height of the photon, the zenith angle, and wavelength. The transmission probability is calculated by the software package MODTRAN \cite{berk2014modtran} which gets a measured atmosphere at the South Pole as input. 
The full telescope simulation includes the response of the telescope optics depending on the wavelength and zenith angle of the incoming photons and the simulation of the silicon photomultiplier (SiPM) response and the DAQ chain. The SiPM response module includes the simulation of the background noise which is mainly produced by the electronic noise and photons of the NSB and cross talk. 
Lastly, the camera images have to be cleaned. An overall threshold of 15$\,$PE will cut away most of the noise signals, but pixels with at least two neighboring pixels above 15$\,$PE will be kept if their peak height is higher than 10$\,$PE and the peak times are within 5$\,$ns. 
An event overall will be removed if it has less than 3 surviving pixels, or
if the ratio of the sum of charges of the inner pixels divided by the sum of charges of the outer pixels is larger than one, with inner/outer referring to the position on the camera board. This cleans out the showers which are not mostly contained in the camera. 

\section{Graph Neural Network reconstruction of air shower properties}\label{secgnn}
A GNN has been implemented to reconstruct the air shower parameters. A GNN in general consists of nodes that contain the event information and edges which define the relationship between the nodes. Each graph can have a different number of nodes and relationships between the nodes, but the structure is constant between events. 
The GNN has been chosen as machine learning method since it can handle a different number of nodes for each event and it is easily expandable to more telescopes and other detector types in future analysis. 
At this point, the GNN uses just a single telescope for the reconstruction and reconstructs all values in relation to the telescope position. 
Five of the MC datasets (neon is kept for testing only) are used for training and validation of the GNN, with 50$\%$ used for training and the remaining percentage split equally into a validation and test dataset. 
\begin{figure}[htb]
	\centering
	\begin{subfigure}[b]{0.48\textwidth}
		\centering
		\includegraphics[width=\textwidth]
  {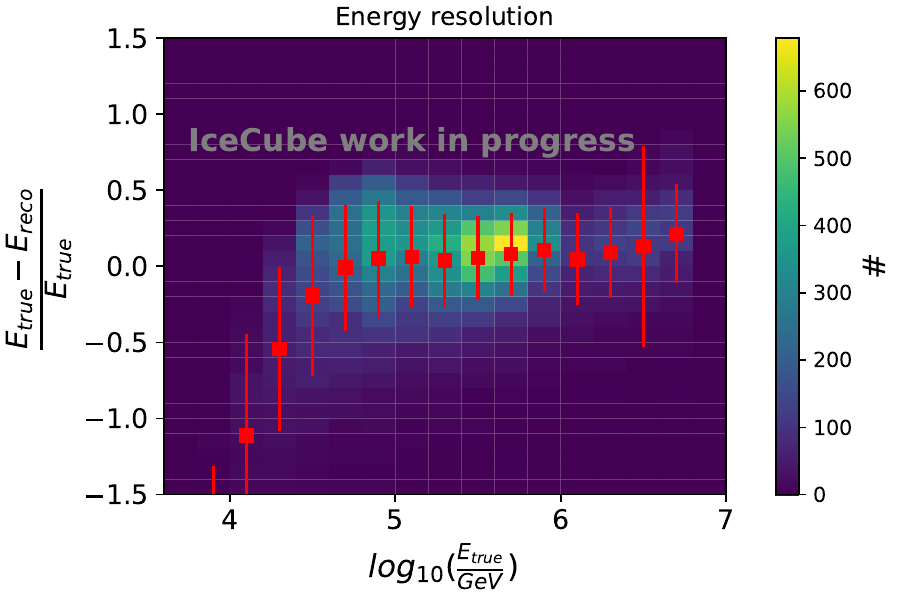}
	\end{subfigure}
	\hfill
	\begin{subfigure}[b]{0.48\textwidth}
		\centering		
		\includegraphics[width=\textwidth]{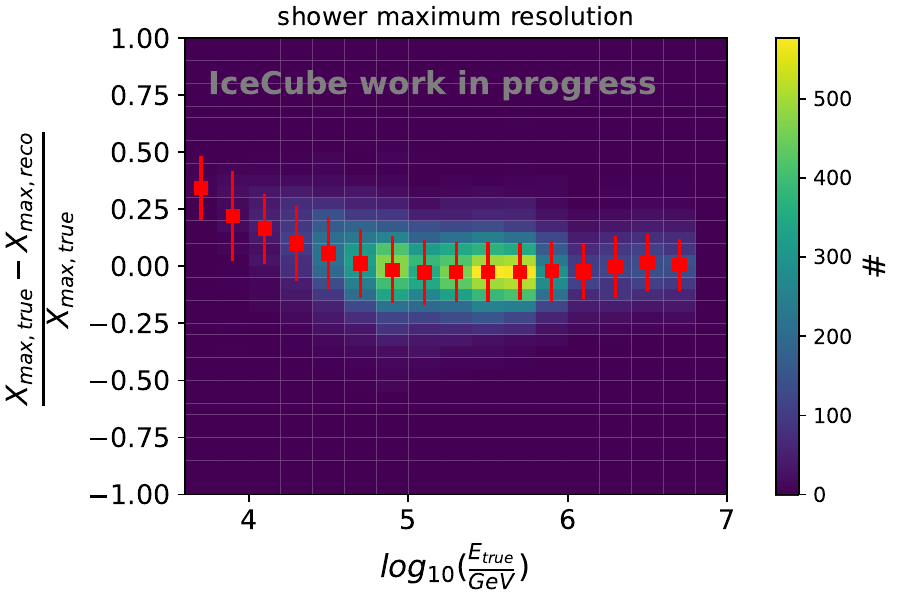}
	\end{subfigure}
	\caption{\label{Plt:GNNresolution} Left: The average energy reconstruction between $10^{4.5}\,$GeV and $10^{6.5}\,$GeV shows a small constant bias of about 5$\%$ and has a resolution of about 50$\%$. Right: The resolution for the shower maximum reconstruction in this energy range has a small constant bias of about 2$\%$ and resolution better than 20$\%$. The red dots are the mean values with the corresponding standard deviation.}
\end{figure}
Input parameters into the GNN nodes are the height and time of the signal peak in a pixel, and the x position and y position of the pixel. The edges of the GNN are constructed in such a way that each pixel knows itself and its neighboring pixel. The GNN structure consists of two convolutional layers which are followed by a pooling layer and 4 dense layers. It reconstructs the following output variables: the cosmic ray primary energy \logEGeV{}, 
the position of the shower core 
(with respect to the position of the telescope, expressed as a distance and an angle), 
the track direction, 
and the slant depth of the shower maximum ($X_{max}$). 
To assist the GNN, some variables (the primary energy, the shower maximum, and the distance) are normalized to be centered on zero, by subtraction of the mean of the training data set, and dividing by the standard deviation. 
Additionally, angular variables (the core position angle, and track zenith and azimuth) are treated in cosine or sine form so as to avoid cyclical behavior and to be between -1 and 1.
Note that the GNN is constructed to reconstruct the total energy of the primary particle; the energy visible in the Cherenkov photons comes from just the electromagnetic part of the air shower. 

The resolution of the energy 
(as measured in \logEGeV) and shower maximum reconstruction are shown in Figure~\ref{Plt:GNNresolution}. Below 4.5, the statistic is low because the telescope is not yet sensitive to all nuclei. Above 6.0, the statistic is lower because the relative amount of simulated events is lower. Above 6.5, the energy reconstruction gets biased because there are no simulated events above 6.75, so the GNN tends to reconstruct lower energy because it seems more probable. On average the energy reconstruction between 4.5 and 6.5 has a small constant bias of about 5$\%$ and a resolution of about 50$\%$. 
Thus, this energy region has been chosen as the analysis range for the IceAct-only composition analysis. The resolution for the shower maximum reconstruction in this energy range has just a small constant bias of about 2$\%$ and is better than 20$\%$. The shower core position can be reconstructed within 50$\,$m, and the angular reconstruction is better than 1$^{\circ}$ for this energy region. This development stage of the GNN shows promising first results for reconstructing the air shower properties with possibilities for future improvements.
\section{Mass sensitive input parameters into to the Random Forest Regressors}\label{secinput}
\begin{figure*}[h]
    \begin{minipage}[t]{0.7\linewidth}
        \begin{subfigure}{0.99\linewidth}
            \centering
            \includegraphics[width=\linewidth]{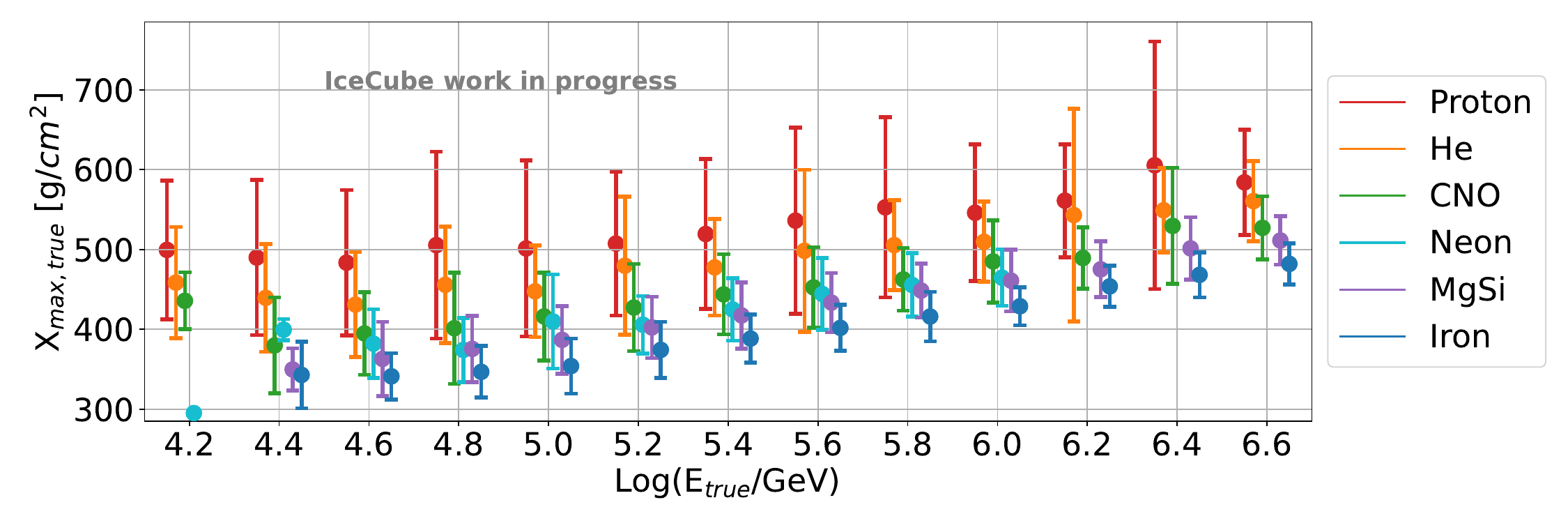}
    
        \end{subfigure}
        \hfill
        \begin{subfigure}{0.99\linewidth}
            \centering
            \includegraphics[width=\linewidth]{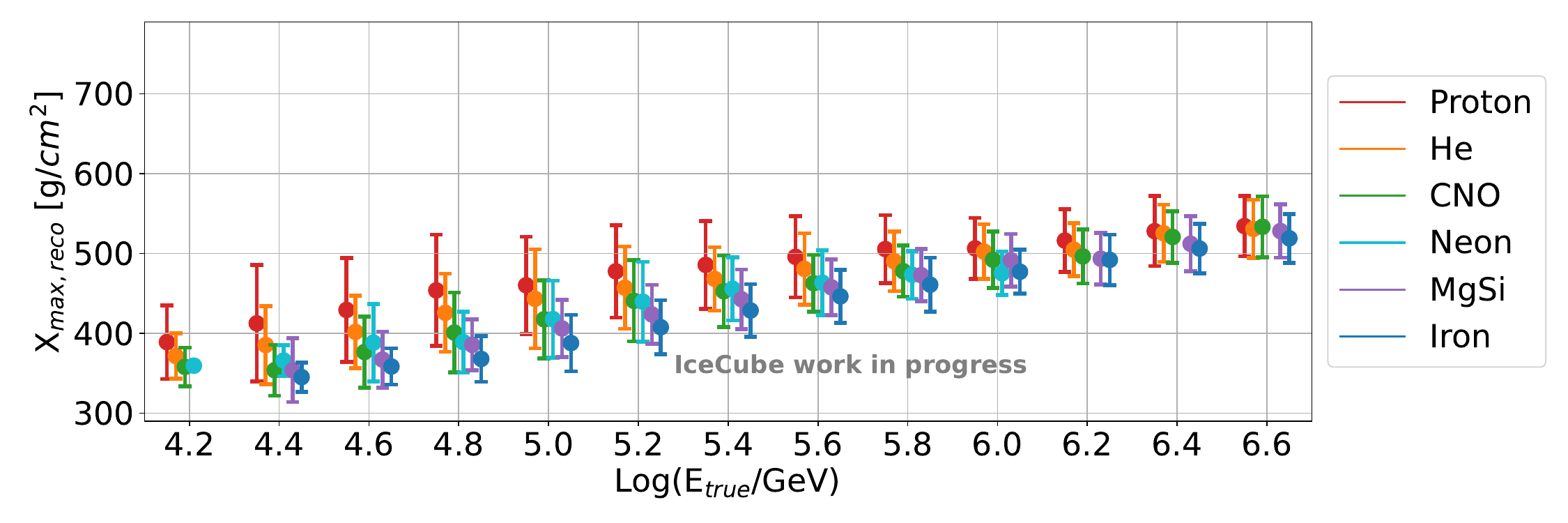}
        \end{subfigure}
    \end{minipage}
    \hfill
    \begin{minipage}{0.3\linewidth}
        \caption{\label{Plt:RFinputGhmax} Top: Position of the shower maximum determined using the true particle distribution vs.~true primary particle energy. Bottom: GNN reconstructed shower maximum position vs.~true primary particle energy. }
    \end{minipage}
\end{figure*}

Two of the most important mass composition sensitive parameters in this analysis are the position of the shower maximum and the number of muons in the deep ice detector. The position of $X_{max}$ is directly reconstructed with the GNN. 
Figure~\ref{Plt:RFinputGhmax} demonstrates the ability to distinguish masses using $X_{max}$: from the true value using the CORSIKA particle distributions, as if with an ideal detector (upper plot), and from the reconstructed value from the GNN (lower plot).
Overall the tendency of lighter nuclei to have a higher $X_{max}$ is reconstructed well. 
As expected, detector resolution and reconstruction lessen the separability when GNN-reconstructed quantities are used instead of true values. The difference between the two plots shows how much room for improvement is possible for the input parameters into the composition analysis. 
The separability also gets visibly smaller above $10^6\,$GeV, probably because the number of events in that region is significantly lower and the GNN has less opportunity to train and validate on outliers.
\begin{figure*}[h]
    \centering
    \begin{minipage}[t]{0.7\linewidth}
        \begin{subfigure}{0.99\linewidth}
            \centering
            \includegraphics[width=\linewidth]{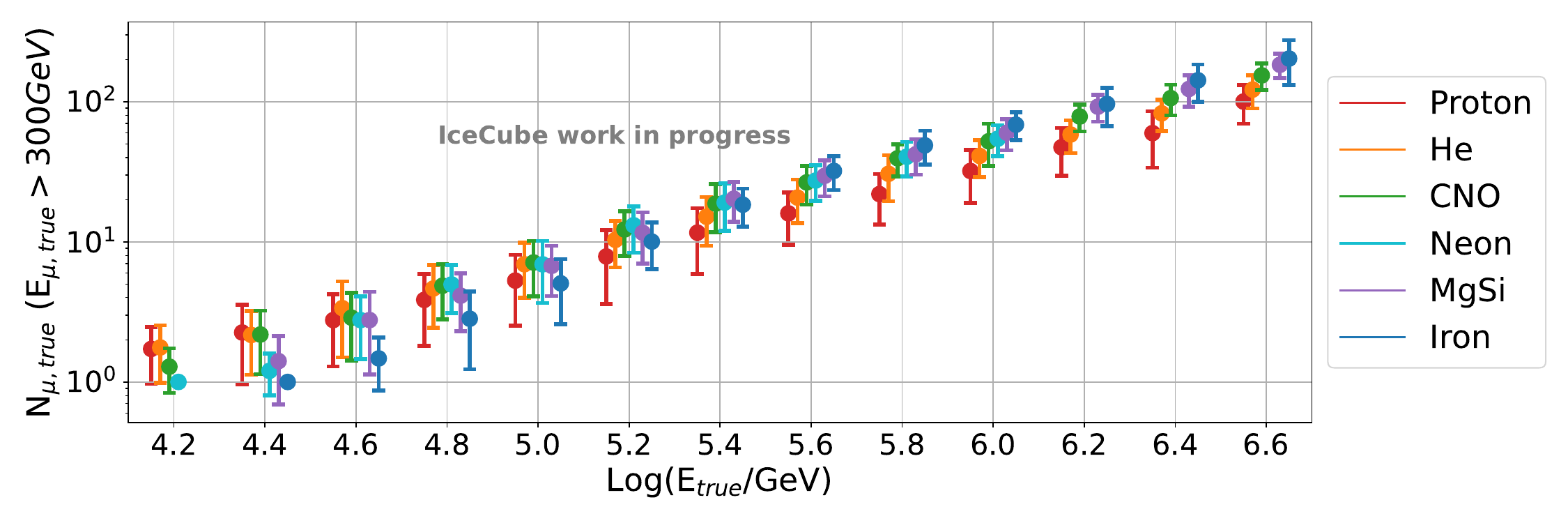}
    
        \end{subfigure}
        \hfill
        \begin{subfigure}{0.99\linewidth}
            \centering
            \includegraphics[width=\linewidth]{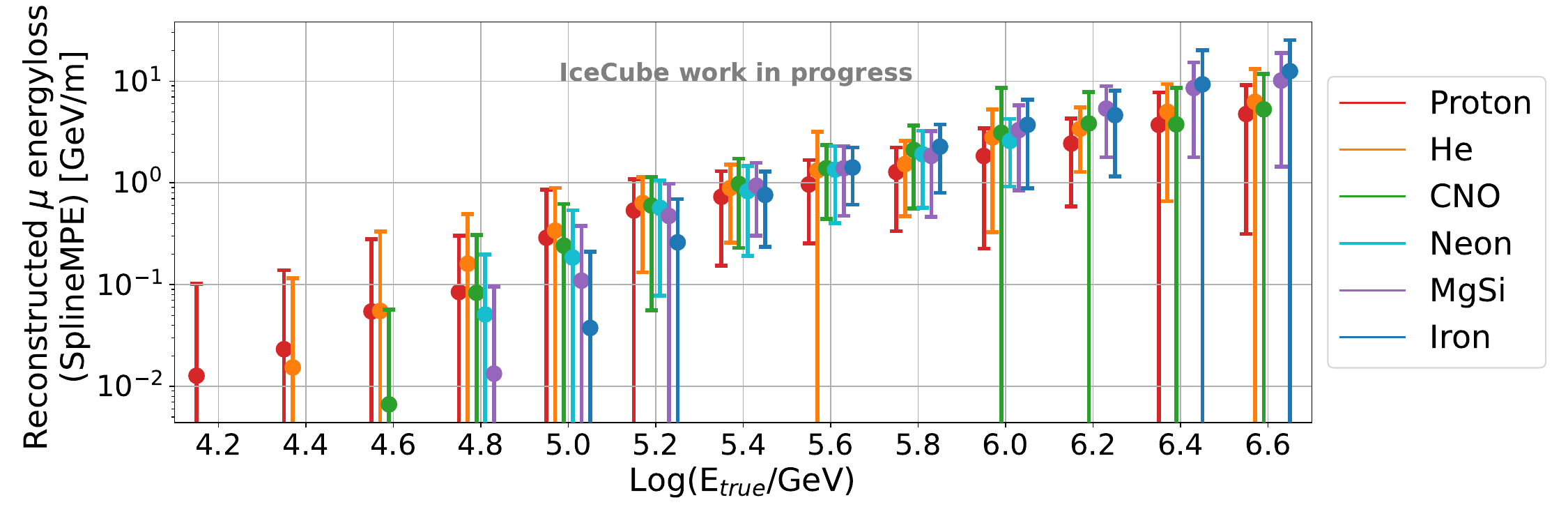}
        \end{subfigure}
    \end{minipage}
    \hfill
    \begin{minipage}{0.28\linewidth}
        \caption{\label{Plt:RFinputMuon} Top: True number of muons above 300$\,$GeV vs.~true primary particle energy. Bottom: Reconstructed muon energy loss vs.~true primary particle energy. Overall, above $\sim$10$^5\,$GeV lighter nuclei having fewer muons exhibit a tendency to have a lower energy loss.}
    \end{minipage}
\end{figure*}

Additionally, the energy loss of the muons in the in-ice detector is used as a proxy for the number of high-energy muons in the air shower. 
Similar effects to the slant depth of the shower maximum can be seen for the separability of the true number of muons 
for each type of nucleus shown in Figure \ref{Plt:RFinputMuon}. The separability is greater for the true number of muons than the separability of the reconstructed deposited energy. The reconstruction of the deposited energy is also not fully efficient for the heavy nuclei for energies below $10^{5.2}\,$GeV, therefore, the composition analysis including IceAct and IceCube is restricted to energies between $10^{5.2}\,$GeV and $10^{6.5}\,$GeV.


\section{Mass composition analysis using Random Forest Regressors}\label{secrf}
\begin{figure*}[h]
    \centering
    \begin{minipage}[t]{0.62\linewidth}
        \begin{subfigure}{0.99\linewidth}
            \centering
            \includegraphics[width=\linewidth]{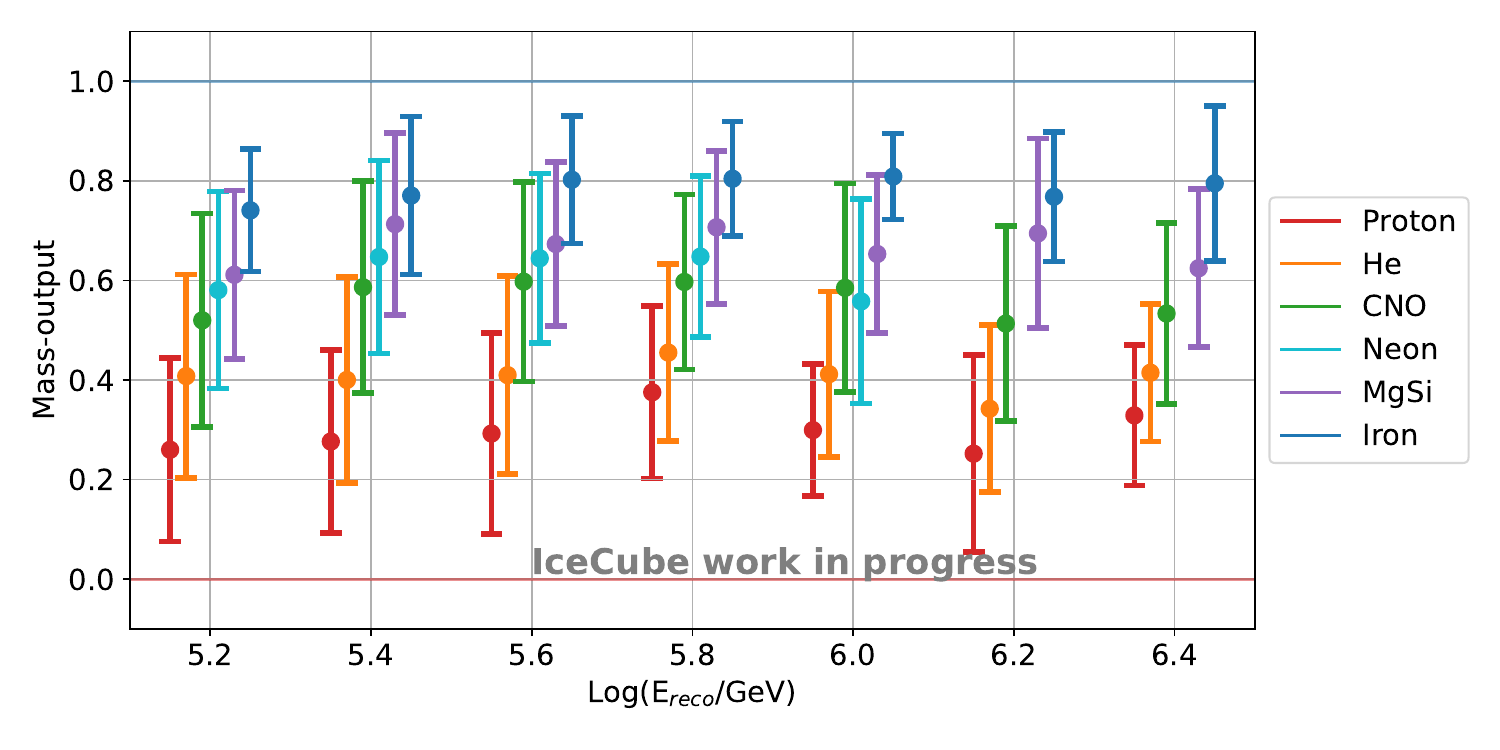}
    
        \end{subfigure}
        \hfill
        \begin{subfigure}{0.99\linewidth}
            \centering
            \includegraphics[width=\linewidth]{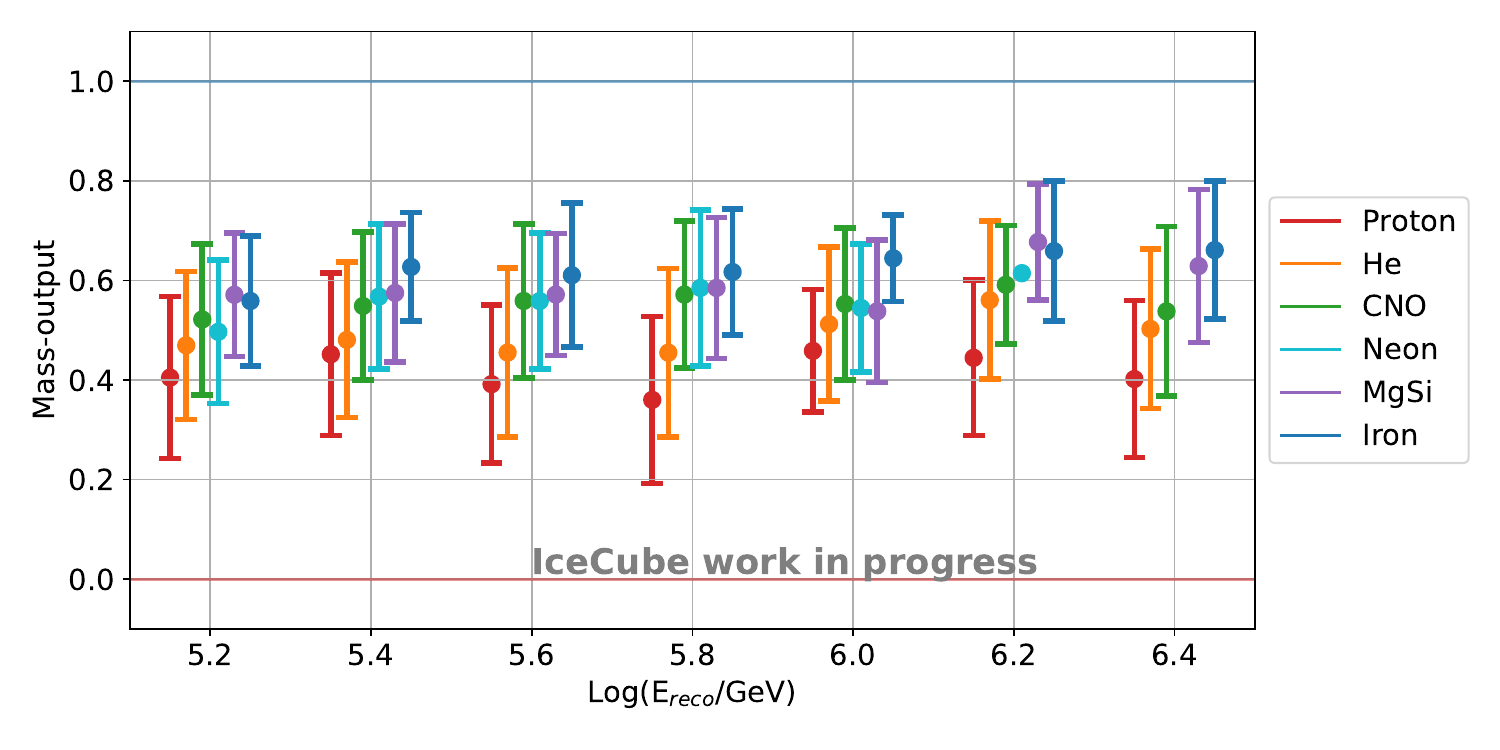}
        \end{subfigure}
    \end{minipage}
    \hfill
    \begin{minipage}{0.33\linewidth}
        \caption{\label{Plt:RFoutput} Top: Mass output vs.~reconstructed energy of the RF analysis using the true values for the IceAct and IceCube response as input into the RF. Bottom: Mass output vs.~reconstructed energy of the RF using reconstructed values for the IceAct and IceCube response as input into the RF. The horizontal lines show the true mass output for proton in red and iron in blue. The top plot shows the ideal results analysis which  are very hard to reach with a real detector response.}
    \end{minipage}
\end{figure*}
As a next step, RFs are used to turn these composition-sensitive observables -- together with additional observables from IceAct and/or IceCube -- into a cosmic ray mass estimator and to reconstruct the primary particle energy.  Two versions of this have been explored: one using IceAct observables alone, and one using IceAct observables together with IceCube observables.  Both RF's are provided with Cherenkov telescope specific parameters as inputs: the length and width of an ellipse \cite{hillas} fitted to the cleaned 
image of the air shower on the telescope camera
, and the total charge measured with the telescope (called the "image size").  The IceAct-only RF is additionally provided the GNN-reconstructed $X_{max}$, the reconstructed zenith of the incoming primary particle, the distance between the shower core and the telescope, and energy reconstructed with the telescope.  The IceAct-IceCube RF is additionally provided the GNN-reconstructed distance between the shower core and the telescope, the energy reconstructed with the telescope and $X_{max}$, and with IceCube reconstructed cosine zenith of the incoming primary particle and muon energy loss.
Both RFs are trained on a data set containing proton, helium, nitrogen, aluminum, and iron primaries. The neon data set is used for testing the RF results. This work uses just the information obtained from one telescope; the extension to a two-telescope analysis is in preparation. The data set is split in two thirds for training and one third for testing. 
The performance of the RF containing IceAct and IceCube input parameters is shown in Figure~\ref{Plt:RFoutput}, for true (top) and reconstructed (bottom) input values. 
The correct behavior is kept using reconstruced observables, with the average proton (red) determined lighter than the average iron nuclei (blue). This is also true for the IceAct only mass composition analysis, which shows a slightly lower separability for the nuclei than the analysis that includes IceCube parameters. 

\section{Template analysis of mass fractions}\label{sec5}
\begin{figure}[h]
	\centering
	\includegraphics[width=0.88\textwidth]{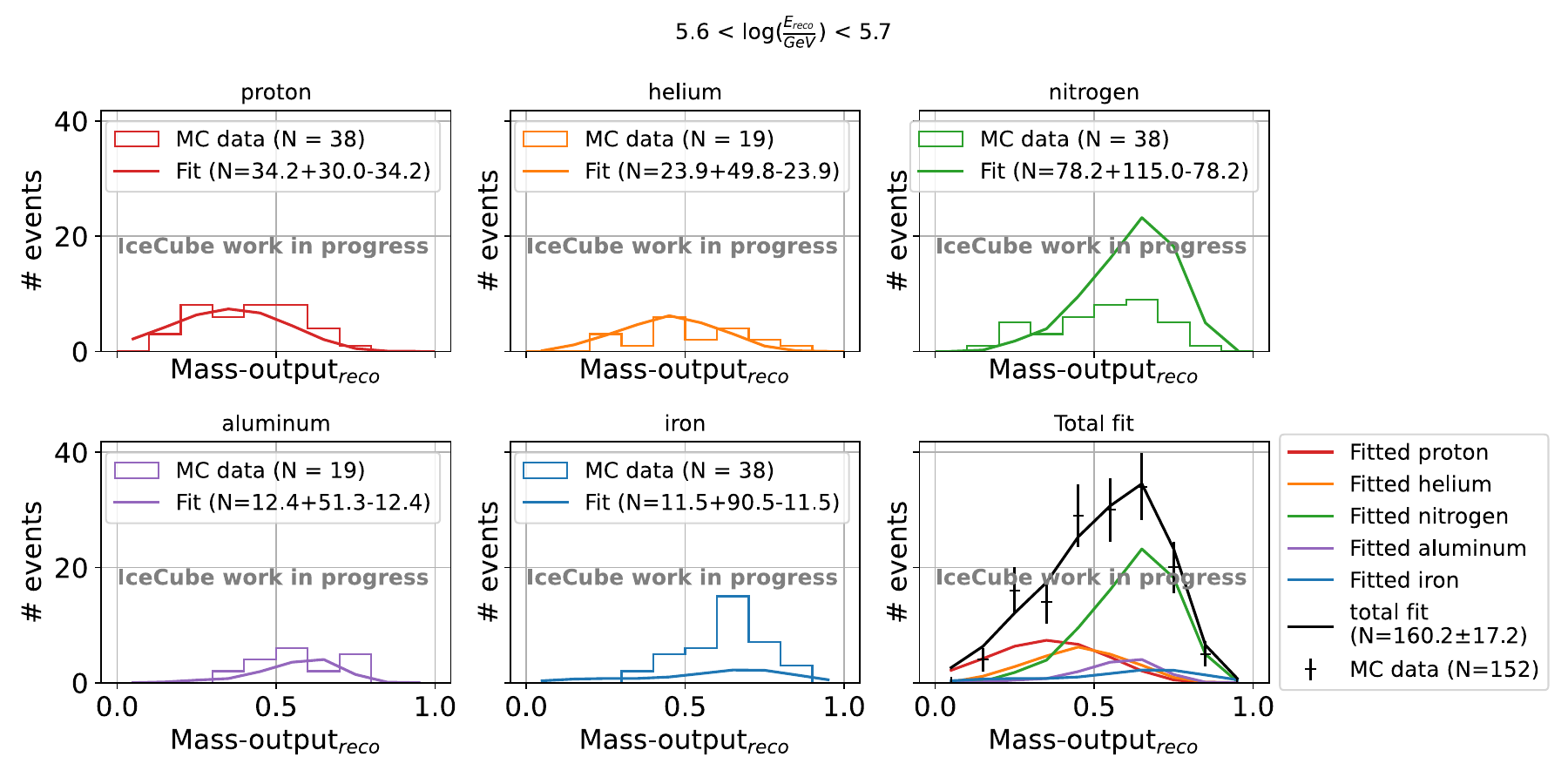}
	\caption{\label{Plt:fittedtemplates}True input into the tested composition hypothesis (histogrammed data) and reconstructed fractions (solid line) for one example energy slice. 
 }
\end{figure}

The mass output is now used to determine the fraction of each nucleus which is present in a data set.
For the template analysis, the remaining data set from testing the RF is split into one third for creating templates that represent the standard output of the RF for each nuclei type and the remaining two thirds of the data for testing different hypotheses of fractions of mass composition. Here, an example composition hypothesis will be constructed to test the overall technique: a mixture in which the fractions of proton, nitrogen, and iron are twice the number of events than the number of events for the helium and aluminum nuclei.  Figure~\ref{Plt:fittedtemplates} shows 
how the template-fitting technique treats this example,
for one energy slice. 
The fit is strictly required to reconstruct the correct number of total events and not less than zero events for any nucleus, which leads to asymmetric error bars. 
The fitted values agree within the errors with the true number of events used in this hypothesis.

\section{Summary and Outlook}\label{secsummary}
Preparation of the data set for performing a mass composition analysis is ongoing and quality parameters to study the atmospheric conditions show promising prospects \cite{Author:2023icrcLars}. Both telescopes have been calibrated and the calibration is stable over the years of operation. 
The GNN reconstruction presented in this work yields resolutions that are a promising first step for future mass composition analyses.
More statistics for the higher energies can increase the reconstruction performance. 
Further improvement might also be achieved by optimizing the GNN parameters.

The RFs can successfully reconstruct the general trend of the mass output for a single telescope analysis for the presented simulated data set. 
Improvements can be achieved by better reconstructed input parameters from the GNN into the RF and more simulated events, especially at high energies. Additionally, the RF settings have not yet been optimized. The template fit method will be used to determine the real fractions for each element group over the whole energy range. 
An example hypothesis has been reconstructed using template fitting. The results show room for improvement achievable by advances in the previous analysis steps. 

A simultaneous reconstruction of two telescopes with one GNN is also in preparation; this should help especially with the energy reconstruction and the reconstruction of the distance between the telescope and the shower core, which are the most difficult to reconstruct with a Cherenkov telescope alone. This should also improve RF analysis where a two-telescope expansion is also in preparation.

\bibliographystyle{ICRC}
\bibliography{references}

%

\clearpage

\section*{Full Author List: IceCube Collaboration}

\scriptsize
\noindent
R. Abbasi$^{17}$,
M. Ackermann$^{63}$,
J. Adams$^{18}$,
S. K. Agarwalla$^{40,\: 64}$,
J. A. Aguilar$^{12}$,
M. Ahlers$^{22}$,
J.M. Alameddine$^{23}$,
N. M. Amin$^{44}$,
K. Andeen$^{42}$,
G. Anton$^{26}$
% ...
\\
\\
% ...
$^{12}$ Universit{\'e} Libre de Bruxelles, Science Faculty CP230, B-1050 Brussels, Belgium \\
% ...
$^{17}$ Department of Physics, Loyola University Chicago, Chicago, IL 60660, USA \\
$^{18}$ Dept. of Physics and Astronomy, University of Canterbury, Private Bag 4800, Christchurch, New Zealand \\
% ...
$^{22}$ Niels Bohr Institute, University of Copenhagen, DK-2100 Copenhagen, Denmark \\
$^{23}$ Dept. of Physics, TU Dortmund University, D-44221 Dortmund, Germany \\
% ...
$^{26}$ Erlangen Centre for Astroparticle Physics, Friedrich-Alexander-Universit{\"a}t Erlangen-N{\"u}rnberg, D-91058 Erlangen, Germany \\
% ...
$^{40}$ Dept. of Physics and Wisconsin IceCube Particle Astrophysics Center, University of Wisconsin{\textendash}Madison, Madison, WI 53706, USA \\
% ...
$^{42}$ Department of Physics, Marquette University, Milwaukee, WI, 53201, USA \\
% ...
$^{44}$ Bartol Research Institute and Dept. of Physics and Astronomy, University of Delaware, Newark, DE 19716, USA \\
% ...
$^{63}$ Deutsches Elektronen-Synchrotron DESY, Platanenallee 6, 15738 Zeuthen, Germany  \\
$^{64}$ Institute of Physics, Sachivalaya Marg, Sainik School Post, Bhubaneswar 751005, India \\
% ...

\subsection*{Acknowledgements}

\noindent
The authors gratefully acknowledge the support from the following agencies and institutions:
USA {\textendash} U.S. National Science Foundation-Office of Polar Programs,
U.S. National Science Foundation-Physics Division,
U.S. National Science Foundation-EPSCoR,
Wisconsin Alumni Research Foundation,
Center for High Throughput Computing (CHTC) at the University of Wisconsin{\textendash}Madison,
Open Science Grid (OSG),
Advanced Cyberinfrastructure Coordination Ecosystem: Services {\&} Support (ACCESS),
Frontera computing project at the Texas Advanced Computing Center,
U.S. Department of Energy-National Energy Research Scientific Computing Center,
Particle astrophysics research computing center at the University of Maryland,
Institute for Cyber-Enabled Research at Michigan State University,
and Astroparticle physics computational facility at Marquette University;
Belgium {\textendash} Funds for Scientific Research (FRS-FNRS and FWO),
FWO Odysseus and Big Science programmes,
and Belgian Federal Science Policy Office (Belspo);
Germany {\textendash} Bundesministerium f{\"u}r Bildung und Forschung (BMBF),
Deutsche Forschungsgemeinschaft (DFG),
Helmholtz Alliance for Astroparticle Physics (HAP),
Initiative and Networking Fund of the Helmholtz Association,
Deutsches Elektronen Synchrotron (DESY),
and High Performance Computing cluster of the RWTH Aachen;
Sweden {\textendash} Swedish Research Council,
Swedish Polar Research Secretariat,
Swedish National Infrastructure for Computing (SNIC),
and Knut and Alice Wallenberg Foundation;
European Union {\textendash} EGI Advanced Computing for research;
Australia {\textendash} Australian Research Council;
Canada {\textendash} Natural Sciences and Engineering Research Council of Canada,
Calcul Qu{\'e}bec, Compute Ontario, Canada Foundation for Innovation, WestGrid, and Compute Canada;
Denmark {\textendash} Villum Fonden, Carlsberg Foundation, and European Commission;
New Zealand {\textendash} Marsden Fund;
Japan {\textendash} Japan Society for Promotion of Science (JSPS)
and Institute for Global Prominent Research (IGPR) of Chiba University;
Korea {\textendash} National Research Foundation of Korea (NRF);
Switzerland {\textendash} Swiss National Science Foundation (SNSF);
United Kingdom {\textendash} Department of Physics, University of Oxford.

\end{document}